\documentclass
[superscriptaddress,secnumarabic,amssymb,amsmath,nobibnotes,aps,prd,showkeys,showpacs,nofootinbib]{revtex4}%
\usepackage{graphicx}
\usepackage{epsf}
\usepackage{bm}
\usepackage{amsmath}
\usepackage{amsfonts}
\usepackage{amssymb}
\usepackage{epstopdf}%
\providecommand{\U}[1]{\protect\rule{.1in}{.1in}}
\newcommand{\be}{\begin{equation}}
\newcommand{\ee}{\end{equation}}

\newcommand{\mincir}{\raise
-3.truept\hbox{\rlap{\hbox{$\sim$}}\raise4.truept\hbox{$<$}\ }}
\newcommand{\magcir}{\raise
-3.truept\hbox{\rlap{\hbox{$\sim$}}\raise4.truept\hbox{$>$}\ }}

\begin{document}
\title{New integrable models and analytical solutions in $f(R)$~cosmology with an
ideal gas}
\author{G. Papagiannopoulos}
\email{yiannis.papayiannopoulos@gmail.com}
\affiliation{Faculty of Physics, Department of Astronomy-Astrophysics-Mechanics University
of Athens, Panepistemiopolis, Athens 157 83, Greece}
\author{Spyros Basilakos}
\email{svasil@academyofathens.gr}
\affiliation{Academy of Athens, Research Center for Astronomy and Applied Mathematics,
Soranou Efesiou 4, 11527, Athens, Greece}
\author{John D. Barrow}
\email{jdb34@hermes.cam.ac.uk}
\affiliation{DAMTP, Centre for Mathematical Sciences, University of Cambridge, Wilberforce
Rd., Cambridge CB3 0WA, UK}
\author{Andronikos Paliathanasis}
\email{anpaliat@phys.uoa.gr}
\affiliation{Instituto de Ciencias F\'{\i}sicas y Matem\'{a}ticas, Universidad Austral de
Chile, Valdivia, Chile}
\affiliation{Institute of Systems Science, Durban University of Technology, PO Box 1334,
Durban 4000, Republic of South Africa}
\keywords{Cosmology; $f(R)$-gravity; Symmetries; Integrability}
\pacs{98.80.-k, 95.35.+d, 95.36.+x}

\begin{abstract}
In the context of $f\left(  R\right)  $-gravity with a spatially flat FLRW
metric containing an ideal fluid, we use the method of invariant
transformations to specify families of models which are integrable. We find
three families of $f(R)$ theories for which new analytical solutions are given
and closed-form solutions are provided.

\end{abstract}
\maketitle


\section{Introduction}

\label{intro}

The discovery of the accelerated expansion of the Universe has led to many
cosmological models which aim to explain this phenomenon, using a spatially
flat geometry and a cosmic dark sector formed by cold dark matter and some
form of dark energy, with negative pressure. Among the large family of
possible cosmological scenarios, the modified gravity models occupy a
much-studied position in cosmological studies, since they provide a way of
explaining the accelerated expansion of the universe, under a modification of
Einstein-Hilbert action.
Traditionally, the simplest term which has been added to the gravitational
action is the squared Ricci scalar which leads to a model of inflation
\cite{tomita, star}. The latter corresponds to a family of models in which the
gravitational action is a function of the Ricci scalar of the underlying
geometry and consequently the cosmological field equations are fourth-order in
time. This class of models belongs to the so called $f\left(  R\right)  $
extended theory of gravity \cite{Buda, BOtt, Bcot} and it has been applied in
various areas of the gravitational and cosmological studies ( see
\cite{Sotiriou,odin1,Teg,Kowal,basnes01,Komatsu,Ade15} and references
therein).

The modification of the Einstein-Hilbert action by introducing other kind of
invariants has lead to an entire menagerie\emph{ }of $f$-theories
\cite{Ferraro,mod1,on2,on3,nes2,mod2,mod3,nes3,Clifton,Clif2,Clif3,Clif4,an4,basnes01,basnes02,pan01,pan02}%
. Some theories are second-order, like general relativity, while others are of
at least 4$^{th}$order. In the context of modified gravity, the field
equations form a system of nonlinear ordinary differential equations which may
not be integrable. In general, the role of integrability in any dynamical
problem is to provide the necessary conditions to compute its solution.
Obviously, the latter achievement is very important in cosmology. For example,
one of the main problems that integrability solves is the determination of the
initial conditions in cosmological simulations. Various methods have been
proposed in order to study the integrability of dynamical systems. In
Liouville's integrability approach to (classical) Hamiltonian systems there is
a point transformation for which the action is determined by the method of
separation of variables. This is equivalent with the existence of a second
conservation law which defines a support manifold such that intersects the
phase space volume. However, this is not the only possibility, since the
Hamilton-Jacobi can be solved explicitly without the existence of a coordinate
system that leads to a separation of variables. In the second case the
conservation law provides a Lie-supported manifold for the dynamical system
\cite{Daskaloy}.

Conservation laws provide sufficient constraints to solve the Hamilton-Jacobi
equation and they are related directly to the existence of transformations
('symmetries')which maintain dynamical invariance. Another way to study the
integrability of a dynamical system is by the method of singularity analysis
in which a necessary condition for success is the existence of a singular
solution. Singularity analysis has been performed in various gravitational
studies, and it can be used to determine important information about the
evolution of the system close to a singularity
\cite{aref3,aref4,aref4a,aref5a,cots1,cots2,aref5,aref6,aref7}.

In this work we are interested in algebraic integrability in the sense that we
will search for those $f\left(  R\right)  $-models for which the polynomial in
the momentum conservation law exists and can be used to write the field
equations as a system of two first-order differential equations. In order to
determine the conservation laws, we use Noether's theorems. Specifically, from
the first theorem we find the necessary conditions that the $f\left(
R\right)  $ theory needs to obey in order for there to exist a (generalized)
symmetry vector, while from the second theorem the conservation law is
determined. This is an analogue of the Ovsiannikov classification of the
nonlinear heat equation \cite{Ovsiannikov} which has been applied in various
gravitational theories and has led to new integrable models (see
\cite{Rosquist,nor0,KotsakisL,nor3,nor0a,nor2,nor5,nor9aa,nor9,darabi,nor10,nor4,terzis2,terzis,christ,nor11}
and references therein). Noether's Theorem is the main mathematical tool that
we use in this study and specifically we select to work within the framework
of the so-called 'contact symmetries' \cite{swartz}. We extend our previous
works \cite{and1,and2} to complete the classification of the integrable models
in $f\left(  R\right)  $-gravity \cite{rr1,rr2}. The plan of the paper is as follows.

In Section \ref{field}, we briefly present the main points of $f\left(
R\right)  $ modified gravity with an ideal gas. Next, we derive the field
equations and we discuss the minisuperspace lagrangian in the context of the
classical Hamiltonian formalism. Sections \ref{newsolutions} and \ref{quint}
then present the main results of our analysis, namely we identify those
families of $f(R)$ models for which the field equations admit extra local
conservation laws and so form integrable dynamical systems.
Finally, our conclusions are presented in Section \ref{con}.

\section{The gravitational field equations}

\label{field} In this section, we introduce the main ingredients of $f(R)$
gravity. Specifically, the modified Einstein-Hilbert action of $f(R)$ gravity
is
\begin{equation}
S=\int dx^{4}\sqrt{-g}\left[  \frac{1}{2k}f\left(  R\right)  +L_{m}\right]
,\label{ac.01}%
\end{equation}
where $R$ is the Ricci scalar, $g$ is the determinant of the metric tensor,
$k\equiv8\pi G$ and $L_{m}$ is the lagrangian function for the matter source.
Varying $S$ with respect to the metric, we obtain the gravitational field
equations
\begin{equation}
f^{\prime}R_{\mu\nu}-\frac{1}{2}fg_{\mu\nu}-\left(  \nabla_{\mu}\nabla_{\nu
}-g_{\mu\nu}\nabla_{\sigma}\nabla^{\sigma}\right)  f^{\prime}=kT_{\mu\nu
},\label{ac.02}%
\end{equation}
where $f^{\prime}\left(  R\right)  =\frac{df}{dR}$, $R_{\mu\nu}$ is the Ricci
tensor, and $T_{\mu\nu}=\frac{\partial L_{m}}{\partial g^{\mu\nu}}$ is the
energy-momentum tensor for the matter source. It is interesting to see that
the field equations (\ref{ac.02}) can be cast in the following form
\begin{equation}
R_{\mu\nu}-\frac{1}{2}Rg_{\mu\nu}=k_{\mathrm{eff}}\left(  T_{\mu\nu}+T_{\mu
\nu}^{f\left(  R\right)  }\right)  ,\label{ac.03}%
\end{equation}
where $k_{\mathrm{eff}}=\frac{k}{f^{\prime}\left(  R\right)  }$. Also, the
quantity $T_{\mu\nu}^{f\left(  R\right)  }$ can be viewed as the effective
energy-momentum tensor of the modifications of the Einstein-Hilbert action,
and is expressed as,
\begin{equation}
T_{\mu\nu}^{f\left(  R\right)  }=\left(  \nabla_{\mu}\nabla_{\nu}-g_{\mu\nu
}\nabla_{\sigma}\nabla^{\sigma}\right)  f^{\prime}+\frac{1}{2}\left(
f-Rf^{\prime}\right)  g_{\mu\nu}.\label{ac.04}%
\end{equation}
Of course, for $f(R)=R$ the above formulas reduce to those of GR.

Now, assuming a spatially flat Friedmann-Lema\^{\i}tre-Robertson-Walker (FLRW)
metric,
\begin{equation}
ds^{2}=-dt^{2}+a^{2}\left(  t\right)  \left(  dx^{2}+dy^{2}+dz^{2}\right)  ,
\label{fr.01}%
\end{equation}
the field equations (\ref{ac.02}) give the modified Friedmann's equations:
\begin{equation}
3f^{\prime}H^{2}=k\rho_{m}+\frac{f^{\prime}R-f}{2}-3Hf^{\prime\prime}\dot{R},
\label{ac.05}%
\end{equation}
and%
\begin{equation}
2f^{\prime}\dot{H}+3f^{\prime}H^{2}=-2Hf^{\prime\prime}\dot{R}-\left(
f^{\prime\prime\prime}\dot{R}^{2}+f^{\prime\prime}\ddot{R}\right)
-\frac{f-Rf^{\prime}}{2}-kp_{m}, \label{ac.06}%
\end{equation}
where $a(t)$ is the scale factor, $H=\dot{a}/a$ is the Hubble
parameter\footnote{Recall that for an arbitrary lapse function $N\left(
t\right)  $ in the line element (\ref{fr.01}) the Hubble function is defined
as $H=\frac{\dot{a}}{Na}$.} and $R=6\left[  \frac{\ddot{a}}{a}+\left(
\frac{\dot{a}}{a}\right)  ^{2}\right]  $ in which the dot denotes derivative
with respect to the comoving proper time, $t$. Notice that $\rho_{m}=T_{\mu
\nu}u^{\mu}u^{\nu}$ and $p_{m}=T_{\mu\nu}\left(  g^{\mu\nu}+u^{\mu}u^{\nu
}\right)  $ are the energy density and pressure of the matter source, where
$u^{\mu}$ is the normalised four-velocity vector. For the equation of state of
the cosmic matter, we use $p_{m}=w_{m}\rho_{m}$ which corresponds to an ideal
gas, namely with$w_{m}=\gamma-1=$ const., and $w_{m}\in\left[  0,1\right]  $
(or $\gamma\in\left[  1,2\right]  $). Utilizing the conservation law
$T_{~~~;\nu}^{\mu\nu}=0$ it is easy to show that $\rho_{m}$ evolves as
$\rho_{m}=\rho_{m0}a^{-3\gamma}$, where $\rho_{m0}$ is the corresponding
density at the present time ($a=a_{0}=1$). It is worth noting that for
dominant relativistic matter we have $\gamma=4/3$ (when $w_{m}=1/3$), while in
the case of pressureless matter we get $\gamma=1$ ( $w_{m}=0$).

If we focus on the first Friedmann equation (\ref{ac.05}), then we can
introduce an effective dark energy sector of (modified) gravitational origin.
Indeed, the dark energy density and pressure are rewritten as
\begin{equation}
\rho_{f}=\frac{f^{\prime}R-f}{2}-3Hf^{\prime\prime}\dot{R}, \label{ac.07}%
\end{equation}%
\begin{equation}
p_{f}=2Hf^{\prime\prime}\dot{R}+\left(  f^{\prime\prime\prime}\dot{R}%
^{2}+f^{\prime\prime}\ddot{R}\right)  +\frac{f-Rf^{\prime}}{2}, \label{ac.08}%
\end{equation}
while the effective equation of state (EoS) parameter $w_{f}=\frac{p_{f}}%
{\rho_{f}}$ is given by \cite{Sotiriou,odin1,nor3},
\begin{equation}
w_{f}=-\frac{\left(  f-Rf^{\prime}\right)  +4Hf^{\prime\prime}\dot{R}+2\left(
f^{\prime\prime\prime}\dot{R}^{2}+f^{\prime\prime}\ddot{R}\right)  }{\left(
f-Rf^{\prime}\right)  +6Hf^{\prime\prime}\dot{R}}. \label{ac.09}%
\end{equation}

\subsection{Minisuperspace Lagrangian}

From the technical point of view it will help our analysis to insert the
Lagrange multiplier \cite{lanm1,lanm2,lanm3} $\lambda$ in the action
(\ref{ac.01}) as follows
\begin{equation}
S=\int dx^{4}\sqrt{-g}\left[  f\left(  R\right)  -\lambda\left[  R-6\left(
\frac{\ddot{a}}{a}+\left(  \frac{\dot{a}}{a}\right)  ^{2}\right)  \right]
+2\rho_{m0}a^{-3\left(  \gamma-1\right)  }\right]  , \label{ac.10a}%
\end{equation}
where we have set $\sqrt{-g}L_{m}=\rho_{m0}a^{-3\left(  \gamma-1\right)  }$
and $k=8\pi G\equiv1$. Now, if we vary the action with respect to $R$
($\frac{\partial S}{\partial R}=0$) we find $\lambda=f^{\prime}\left(
R\right)  $. Therefore, using the latter condition it is straightforward to
obtain the overall Lagrangian of the current dynamical problem, namely
\begin{equation}
L\left(  a,\dot{a},R,\dot{R}\right)  =6af^{\prime}\dot{a}^{2}+6a^{2}%
f^{\prime\prime}\dot{a}\dot{R}+a^{3}\left(  f^{\prime}R-f\right)  +2\rho
_{m0}a^{-3\left(  \gamma-1\right)  }. \label{ac.11}%
\end{equation}
One may check that with the aid of Eq.(\ref{ac.11}) the Euler-Lagrange
equations provide Eq.(\ref{ac.06}) and the definition of the Ricci scalar
respectively. Furthermore, introducing a lapse function $N$ in the FLRW metric
(\ref{fr.01}) the total Lagrangian becomes
\begin{equation}
L\left(  N,a,\dot{a},R,\dot{R}\right)  =\frac{1}{N}\left(  6af^{\prime}\dot
{a}^{2}+6a^{2}f^{\prime\prime}\dot{a}\dot{R}\right)  +Na^{3}\left(  f^{\prime
}R-f\right)  +2\rho_{m0}Na^{-3\left(  \gamma-1\right)  }, \label{ac.12}%
\end{equation}
The fact that the Lagrangian is independent from derivatives of $N$ implies
that the corresponding Hessian vanishes and so the system is singular. Indeed,
the first Friedmann equation (\ref{ac.05}) is the constraint equation, namely
$\frac{\partial L}{\partial N}=0,$ and it plays a central role in the the
Hamiltonian formalism as well as in the quantization of gravity. Moreover,
without loss of generality we assume $N=N\left(  a,R\right)  $ which means
that the constraint equation (\ref{ac.05}) can be viewed as a conservation law
of the field equations. Lastly, we stress that based on the effective
gravitational parameter $k_{\mathrm{eff}}=\frac{k}{f^{\prime}\left(  R\right)
}\equiv\frac{1}{f^{\prime}\left(  R\right)  }$ we can provide the
scalar-tensor representation of the $f\left(  R\right)  $ gravity which is
equivalent to that of Brans-Dicke theory in the form of O'Hanlon
\cite{Hanlon}. In particular, considering an effective scalar field such as
$\phi=f^{\prime}\left(  R\right)  $ the action integral (\ref{ac.10a}) takes
the following form
\begin{equation}
S=\int dx^{4}\sqrt{-g}\left[  \phi R+V\left(  \phi\right)  -2\rho
_{m0}a^{-3(\gamma-1)}\right]  \label{ac.14}%
\end{equation}
with
\begin{equation}
L\left(  N,a,\dot{a},\phi,\dot{\phi}\right)  =\frac{1}{N}\left(  6a\phi\dot
{a}^{2}+6a^{2}\dot{a}\dot{\phi}\right)  +Na^{3}V\left(  \phi\right)
+2\rho_{m0}Na^{-3\left(  \gamma-1\right)  }. \label{ac.15}%
\end{equation}
where we have set $V\left(  \phi\right)  =f^{\prime}R-f$. Obviously, this
result corroborates, in the effective scalar field language, the transit of
the $f(R)$ theory into the scalar-tensor regime which is a second-order theory
\cite{Bcot}.

The scalar field description of $f\left(  R\right)  $ gravity is useful
because the order of the theory is reduced, while at the same time the number
of the dependent variables is increased. Moreover, with the aid of the scalar
field description one can use the Lagrange multiplier towards extracting a
classical point-like Lagrangian, which can be used in order to write the
corresponding Hamiltonian and thus to derive the field equations
\cite{sot1,ferr11}. Therefore, utilizing the latter mathematical treatment we
can apply the known results of analytical mechanics in order to study the
integrability of our dynamical system \cite{arnold}. As an example, it is well
known in analytical mechanics that the Legendre transformation relates the
second-order Euler-Lagrange equations with the first-order Hamilton's
equations, while the solutions and the integrability survive through the
transformation. Following a similar ideology, one may easily understand the
role of the transformation $\phi=f^{\prime}\left(  R\right)  $ in $f\left(
R\right)  $ gravity, namely it decreases the order of the theory from four to
two (\cite{lanm1,lanm2,lanm3,sot1,ferr11}), while the solution trajectories
remain invariant.

In addition, there is a conformal equivalence between $f(R)$ gravity theories
and general relativity with a scalar field. The former theories are 4th order
in the metric variables and the latter are second order in the effective
scalar field, which is determined by the logarithm of the scalar curvature,
$R$. From the point of view of the initial value problem, the general solution
of general relativity plus a scalar field is determined by 6 independently
arbitrary functions of the three spatial variables on a hypersurface of
constant time, whereas the general solution of $~f\left(  R\right)  $ gravity
in vacuum is determined by 16 arbitrary spatial functions \cite{star00,bar00}.
However, if the metric is specialised to the zero curvature Friedmann model
then no free functions are required in either case. Likewise, all vacuum or
trace-free fluid solutions of general relativity are also particular solutions
of$~f\left(  R\right)  ~$gravity theories if $f\left(  0\right)  =0$ and are
characterised by the same number of free spatial functions in both theories.

\subsubsection{Hamilton's equations}

Using the standard Hamiltonian approach, and with the aid of Eq.(\ref{ac.15}),
we calculate the canonical momenta
\begin{equation}
Np_{a}=12a\phi\dot{a}+6a^{2}\dot{\phi}~,~Np_{\phi}=6a^{2}\dot{a},
\label{ac.16}%
\end{equation}
so the Hamiltonian function becomes
\begin{equation}
\mathcal{H}=N\left[  \frac{p_{a}p_{\phi}}{6a^{2}}-\frac{\phi p_{\phi}^{2}%
}{6a^{3}}-a^{3}V\left(  \phi\right)  -2\rho_{m0}a^{-3\left(  \gamma-1\right)
}\right]  , \label{ac.17}%
\end{equation}
where the field equations of the previous section are those of Hamilton's
equations and the constraint $\mathcal{H}=0.$ In order to proceed with the
dynamical analysis we need to introduce the functional form of the lapse
function $N$. Without loss of generality and in order to simplify the field
equations we prefer to utilize $N=a^{3\left(  \gamma-1\right)  }$ and so the
equations of motion reduce to the following system:\emph{ \ \ \ \ \ \ }
\begin{equation}
\dot{a}=\frac{a^{3\gamma-5}}{6}p_{\phi}~~,~\dot{\phi}=\frac{a^{3\gamma-5}}%
{6}\left(  p_{a}-\frac{\phi}{a}p_{\phi}\right)  , \label{ac.18}%
\end{equation}%
\begin{equation}
\dot{p}_{\phi}=\frac{a^{3\gamma-6}}{6}p_{\phi}^{2}+a^{3\gamma}V_{,\phi},
\label{ac.19}%
\end{equation}
and%
\begin{equation}
\dot{p}_{a}=-\frac{a^{3\gamma-7}}{6}p_{\phi}\left(  \left(  3\gamma-5\right)
ap_{a}-3\left(  \gamma-2\right)  \phi p_{\phi}\right)  +3\gamma a^{3\gamma
-1}V\left(  \phi\right)  . \label{ac.20}%
\end{equation}
Dynamically speaking, these form a two-dimensional autonomous Hamiltonian
system. Except for the nominal conservation law, namely $\mathcal{H}=0$, it is
essential to determine a second conservation law in order to characterize the
dynamical system as integrable. However, the existence of a second
conservation law depends strongly on the functional form of the effective
potential $V\left(  \phi\right)  $, or equivalently, on the form of $f\left(
R\right)  $.\ It is important to mention that our analysis and the solutions
that we find below can be transformed for any other lapse function $N\left(
t\right)  $, for more details we refer the reader to the discussion in
\cite{an00}.

In the rest of the paper we will study a family of effective potentials for
which the corresponding field equations are integrable.

\section{Analytic Solutions}

\label{newsolutions} The fact that the field equations in $f\left(  R\right)
$ gravity form a canonical Hamiltonian system implies that we can use the
basic tools of classical mechanics in order to investigate the integrability
of the system. Specifically, we are interested in those cases where the field
equations are invariant under generalized symmetries and are linear in the
momentum: the so-called contact symmetries. These symmetries provide quadratic
conservation laws in the momentum and they can be determined by using
Noether's second theorem. In Appendix \ref{LBsym} we provide the basic
mathematical tools of the method used here. More details regarding the
application of generalized symmetries in modified theories of gravity can be
found in refs. \cite{and1,and2,rr2}.

The application of the generalized symmetries provides the following general
form of the effective potential
\begin{equation}
V\left(  \phi\right)  =V_{1}\phi^{P}+V_{2}\phi^{Q}, \label{ac.21}%
\end{equation}
where the constants $P,Q$ are functions of the barotropic parameter $\gamma$.
Obviously, this potential describes a large body of $f(R)$ models. In what
follows, we discuss the different cases for the parameters $P,Q~$,which follow
from the symmetry conditions in order for the field equations to admit extra
conservation laws.

\subsection{Model A}

For $(P,Q)=(0,-\frac{1}{2})$ the potential is
\begin{equation}
V_{A}(\phi)=V_{1}+V_{2}\phi^{-\frac{1}{2}}, \label{ac.22}%
\end{equation}
and the field equations (\ref{ac.18})-(\ref{ac.20}) admit the following extra
conservation law:%
\begin{equation}
I_{A}=6a^{9-6\gamma}\dot{a}\left[  \left(  3\gamma-4\right)  \phi\dot{a}%
+a\dot{\phi}\right]  -a^{5}\left(  \frac{3}{5}V_{1}\gamma+V_{2}\phi^{-\frac
{1}{2}}\right)  , \label{ac.23}%
\end{equation}
for $\gamma\neq\frac{5}{3}$, or
\begin{equation}
\bar{I}_{A}=30a^{-1}\dot{a}\left[  \phi\left(  \ln a-1\right)  \dot{a}%
+\dot{\phi}a\ln a\right]  -5a^{5}\ln a\left(  V_{1}+V_{2}\phi^{-\frac{1}{2}%
}\right)  +V_{1}a^{5} \label{ac.23b}%
\end{equation}
for $\gamma=5/3$. At this point we should mention that the above effective
potential (\ref{ac.22}) has been studied in \cite{rr2} and it corresponds to
$f\left(  R\right)  \simeq R^{\frac{1}{3}}-f_{0}$ gravity. However, the
conservation law (\ref{ac.23}) is more general with respect to that of
\cite{rr2}, since in the latter article only the case of a dust fluid
($\gamma=1$) has been used.

Performing the transformation
$a=x^{-\frac{1}{3\gamma-5}}~,~\phi=yx^{\frac{1}{3\gamma-5}}$, for $\gamma
\neq\frac{5}{3}$ the constraint equation and the conservation law $I_{A}$ are
written as,
\begin{equation}
\frac{\left(  5-3\gamma\right)  }{6}p_{x}p_{y}-V_{1}x^{\frac{3\gamma
}{5-3\gamma}}-V_{2}x^{\frac{6\gamma+1}{2\left(  5-3\gamma\right)  }}%
y^{-\frac{1}{2}}-2\rho_{m0}=0, \label{ac.24}%
\end{equation}
and%
\begin{equation}
\frac{\left(  5-3\gamma\right)  }{6}p_{y}\left(  yp_{y}-xp_{x}\right)
+\frac{3}{5}\gamma V_{1}x^{\frac{5}{5-3\gamma}}+V_{2}x^{\frac{11}{2\left(
5-3\gamma\right)  }}y^{-\frac{1}{2}}-I_{A}=0, \label{ac.25}%
\end{equation}
with
\begin{equation}
\dot{x}=\frac{5-3\gamma}{6}p_{y}~,~\dot{y}=\frac{5-3\gamma}{6}p_{x}.
\label{ac.26}%
\end{equation}
Hence, from Eqs.(\ref{ac.24}), (\ref{ac.25}), the action $S\left(  x,y\right)
$ can be determined and the field equations are reduced to those of
(\ref{ac.26}) where%
\begin{equation}
p_{x}=\frac{\partial S\left(  x,y\right)  }{\partial x}\text{ and~\ }%
p_{y}=\frac{\partial S\left(  x,y\right)  }{\partial y}\,. \label{ac.27}%
\end{equation}
We find after some calculations that in the cases of dust ($\gamma=1$) and
relativistic matter ($\gamma=4/3$) the corresponding actions $S\left(
x,y\right)  $ are given by
\begin{equation}
S\left(  x,y\right)  =-\frac{2\sqrt{15}}{5}\sqrt{y\left(  5I_{A}+10\rho
_{m0}x+2V_{1}x^{\frac{5}{2}}\right)  }-\int\frac{\sqrt{15}V_{2}x^{\frac{7}{4}%
}dx}{\sqrt{\left(  5I_{A}+10\rho_{m0}x+2V_{1}x^{\frac{5}{2}}\right)  }%
}~\text{for }\gamma=1 \label{ac.28}%
\end{equation}
and%
\begin{equation}
S\left(  x,y\right)  =-\frac{2\sqrt{30}}{5}\sqrt{y\left(  5I_{A}+10\rho
_{m0}x+V_{1}x^{5}\right)  }-\int\frac{\sqrt{30}V_{2}x^{\frac{9}{2}}dx}%
{\sqrt{\left(  5I_{A}+10\rho_{m0}x+V_{1}x^{5}\right)  }}~,~\text{for }%
\gamma=4/3. \label{ac.29}%
\end{equation}

\subsubsection{Special case}

In order to complete the analysis of the potential (\ref{ac.22}) we consider
the case of $\gamma=\frac{5}{3}$. Utilizing the coordinate transformation
~$a=e^{u},~\phi=ve^{-u}$ the constraint equation and the conservation law
$\bar{I}_{A}$ become
\begin{equation}
\frac{p_{u}p_{v}}{6}-V_{1}e^{5u}-e^{\frac{11}{2}u}v^{-\frac{1}{2}}-2\rho
_{m0}=0, \label{ac.30}%
\end{equation}
\qquad and%
\begin{equation}
\frac{5}{6}p_{v}\left(  up_{u}-vp_{v}\right)  +V_{1}e^{5u}\left(  1-5u\right)
-5V_{2}e^{\frac{11}{2}u}uv^{-\frac{1}{2}}-\bar{I}_{A}=0. \label{ac.31}%
\end{equation}
Therefore, the action is calculated to be%
\begin{equation}
S\left(  u,v\right)  =-\frac{2\sqrt{30}}{5}\sqrt{v\left(  10u\rho_{m0}%
+V_{1}e^{5u}-\bar{I}_{A}\right)  }-\sqrt{30}V_{2}\int\frac{e^{\frac{11}{2}u}%
}{\sqrt{\left(  10u\rho_{m0}+V_{1}e^{5u}-\bar{I}_{A}\right)  }}du
\label{ac.32}%
\end{equation}
while the field equations are reduced to the 2-dimensional system,%
\begin{align}
\dot{u}  &  =-\frac{\sqrt{30}}{v}\sqrt{\left(  10u\rho_{m0}+V_{1}e^{5u}%
-\bar{I}_{A}\right)  },\label{ac.33}\\
\dot{v}  &  =\frac{\sqrt{30}}{6}\frac{V_{1}v+2\rho_{m0}v+\sqrt{v}V_{2}%
e^{\frac{11}{2}u}}{\sqrt{\left(  10u\rho_{m0}+V_{1}e^{5u}-\bar{I}_{A}\right)
v}}. \label{ac.34}%
\end{align}

For large values of $u$, that is in the late universe, this system
approximates to
\begin{equation}
\dot{u}\simeq-\frac{\sqrt{30}}{v}\sqrt{\left(  V_{1}e^{5u}\right)  }~,~\dot
{v}\simeq\frac{\sqrt{30}}{6\sqrt{V_{1}}}V_{2}e^{3u},
\end{equation}
which means that $\int\frac{dv}{v}=-\left(  \frac{V_{2}}{6V_{1}}\right)  \int
e^{\frac{1}{2}u}du$ and ~$\ln\left(  v\right)  =\left(  -\frac{V_{2}}{3V_{1}%
}\right)  e^{\frac{1}{2}u}~$. In this case, after some calculations, we find
\begin{equation}
\phi\left(  a\right)  \simeq\ -\frac{1}{9}\frac{V_{2}}{V_{1}}a^{\frac{5}{2}%
}+\phi_{0}.
\end{equation}
Lastly, the Hubble function for large values of~$a$ takes the form
\begin{equation}
H\simeq a^{-\frac{1}{4}}\exp\left(  \frac{V_{1}V_{2}}{3}\sqrt{a}\right)  .
\end{equation}
Also, the effective equation of state (EoS) parameter is found to be
\begin{equation}
w_{\mathrm{eff}}\left(  a\right)  =-1-\frac{\sqrt{30V_{1}}}{18}\left(
3-2V_{2}V_{1}\sqrt{a}\right)  a^{\frac{11}{4}}. \label{sss1}%
\end{equation}

\begin{figure}[t]
\centering\includegraphics[scale=0.45]{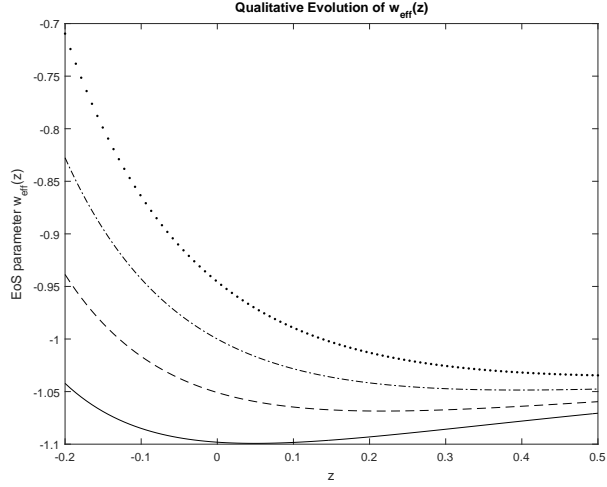}
\caption{Qualitative evolution of the equation of state parameter
(\ref{sss1}). For the plot we selected $V_{2}=2$. The solid line is for
$V_{1}=0.65,~$the dash-dash line for $V_{1}=0.70$ while the dash-dot and
dot-dot liens are for $V_{1}=0.75$ and $V_{1}=0.80$ respectively.}%
\label{fig1}%
\end{figure}

Notice that, in fig. \ref{fig1}, we present the redshift evolution of the
latter EoS parameter.

\subsection{Model B}

In this class of models the constants $P~$and~$Q$ in the effective potential
(\ref{ac.21}) have the following simple expressions%

\begin{equation}
P=-\frac{\gamma}{\gamma-2}~~,~Q=-\frac{3\gamma-10}{3\gamma-4}. \label{ac.35}%
\end{equation}
Obviously, the prohibited values $\gamma\neq4/3$ and $\gamma\neq2$ imply that
in this case the cosmic fluid cannot be that of radiation or stiff matter.
These two special cases need to be studied separately.

Now, the conservation law is
\begin{equation}
I_{B}=a^{10-6\gamma}\left[  \left(  3\gamma-4\right)  \phi\dot{a}+a\dot{\phi
}\right]  ^{2}-\frac{2\left(  3\gamma-5\right)  }{3}V_{2}a^{6}\phi^{\frac
{6}{3\gamma-4}}. \label{ac.36}%
\end{equation}
for $\gamma\neq\frac{5}{3}$. At this point we use the normal coordinates
$\left(  x,y\right)  =\left(  re^{\theta},re^{-\theta}\right)  $, where the
pair $\left(  x,y\right)  $ is given by $a=x^{-\frac{1}{3\gamma-5}}%
~,~\phi=yx^{\frac{1}{3\gamma-5}}$, for $\gamma\neq\frac{5}{3}$. It is worth
noting that in the new coordinates the conservation law is that of the Lewis
invariant which means that the dynamical system admits as a conservation
law\footnote{For more applications of the Ermakov-Pinney system in
gravitational theories and cosmology see \cite{erm0,erm3,erm1,erm2}.} the
Lewis-invariant \cite{LewisI}. Within this framework, we define the constraint
equation and the conservation law, $I_{B}$, by%

\begin{equation}
\left(  3\gamma-5\right)  \left(  \frac{p_{r}^{2}}{24}+\frac{p_{\theta}^{2}%
}{24r^{2}}\right)  -V_{1}r^{-\frac{2\gamma}{\gamma-2}}-\frac{V_{2}}{r^{2}%
}e^{\frac{12\theta}{4-3\gamma}}-2\rho_{m0}=0,
\end{equation}

\begin{equation}
\ \ \frac{p_{\theta}^{2}}{36}+2V_{2}\left(  \frac{5}{3}-\gamma\right)
e^{\frac{12\theta}{4-3\gamma}}-I_{B}=0,
\end{equation}
with
\begin{equation}
\dot{r}=\frac{5-3\gamma}{12}p_{r}\,~,~\dot{\theta}=-\frac{3\gamma-5}{12r^{2}%
}p_{\theta}.
\end{equation}

After some simple algebra, we obtain the corresponding action $S\left(
r,\theta\right)  $%

\begin{align}
S\left(  r,\theta\right)   &  =\int\sqrt{72\left(  \left(  \gamma-\frac{5}%
{3}\right)  V_{2}e^{\frac{-12\theta}{-4+3\gamma}}-\bar{I}_{B}\right)  }%
d\theta+\\
&  -\ \int\sqrt{72\left(  -2\rho_{m0}(\gamma-\frac{5}{3})-(\gamma-\frac{5}%
{3})V_{1}r^{-\frac{2\gamma}{-2+\gamma}}-\frac{1}{72}\bar{I}_{B}r^{-2}\right)
}dr.
\end{align}
Using the conservation law, we find that the constraint equation reads
\begin{equation}
\frac{p_{r}^{2}}{24}-\left(  3\gamma-5\right)  V_{1}r^{-\frac{2\gamma}%
{\gamma-2}}+\frac{3I_{B}}{r^{2}}=2\left(  3\gamma-5\right)  \rho_{m0}.
\label{acc}%
\end{equation}
Inserting the expression $\dot{r}=\frac{5-3\gamma}{12}p_{r}$ into
Eq.(\ref{acc}), we have
\[
\frac{\left(  3\gamma-5\right)  \dot{r}^{2}}{3456}-V_{1}r^{-\frac{2\gamma
}{\gamma-2}}+\frac{3I_{B}}{\left(  3\gamma-5\right)  r^{2}}=2\rho_{m0}.
\]

Interestingly, in the case of $\gamma=0$ ($w_{m}=-1$), the above equation
becomes%
\begin{equation}
\frac{5}{3456}\dot{r}^{2}\ +\frac{3I_{B}}{5}r^{-2}=V_{1}-2\rho_{m0}%
=C(\rho_{m0},V_{1}).
\end{equation}
a solution of which is
\[
r^{2}\left(  t\right)  =\frac{3}{5C}(c+I_{B})+\frac{3}{5}(-c_{1}\bar{C}%
t+\bar{C}t^{2})=r_{0}+\frac{3}{5}(-c_{1}\bar{C}t+\bar{C}t^{2})
\]
where $c_{1}$ is an integration constant, $C$ is a constant that depends from
$(\rho_{m0},V_{1})$ and $\bar{C}=1152C$. Hence, this analytical solution
behaves as $~r\simeq t$ and $r\simeq t^{2}$ for small and large values of $t,$ respectively.

Moreover, in the case of $\bar{I}_{B}=0$, it follows that~$e^{\theta\left(
t\right)  }=\Theta_{0}\left(  \int\frac{dt}{r\left(  t\right)  ^{2}}\right)
^{\lambda\left(  \gamma\right)  },$ which implies that for small or large
values of $t$ we have~$a\left(  t\right)  \simeq t^{\lambda_{1}\left(
\gamma\right)  }\left(  \ln t\right)  ^{\lambda_{2}\left(  \gamma\right)  }%
~$and~$a\left(  t\right)  $ $\simeq t^{\lambda_{1}\left(  \gamma\right)
}e^{t^{\lambda_{3}\left(  \gamma\right)  }}$, where $\lambda\left(
\gamma\right)  $ are constants related to the barotropic index $\gamma.$
Recall that these solutions are expressed in the time in which the lapse
function is $N\left(  t\right)  =a^{3\left(  \gamma-1\right)  }.$

Below, we will consider some special cases, namely $\gamma=5/3$, $4/3$ and $2$.

\subsubsection{Special case $\gamma=5/3$}

Here we consider $\gamma=\frac{5}{3}$, hence from Eq.(\ref{ac.35}) we have
$P=Q=5$ and the effective potential is given by $V(\phi)\propto\phi^{5}$. The
coordinates that we use here are those of section 3.1.1, namely $\left(
u,v\right)  $. In this context, we obtain the constraint equation and the
conservation law $I_{B}:$%

\begin{equation}
\frac{p_{u}p_{v}}{6}+(V_{1}+V_{2})v^{5}+2\rho_{m0}=0,
\end{equation}
$\ $%
\begin{equation}
\frac{p_{u}^{2}}{36}-I_{B}=0.
\end{equation}
with $\dot{v}=\frac{1}{6}p_{u}~,~\dot{u}=\frac{1}{6}p_{v}$. Here, the action
takes the form
\begin{equation}
S\left(  u,v\right)  =6\sqrt{I_{B}}u+\frac{2\left\vert \rho_{m0}\right\vert
}{\sqrt{I_{B}}}v-\frac{\left(  V_{1}+V_{2}\right)  }{36\sqrt{I_{B}}}v^{6};
\end{equation}
that is,%
\begin{equation}
\dot{v}=\sqrt{I_{B}}~\text{and }\dot{u}=\frac{2\left\vert \rho_{m0}\right\vert
}{6\sqrt{I_{B}}}-\frac{\left(  V_{1}+V_{2}\right)  }{36\sqrt{I_{B}}}v^{5},
\end{equation}
or equivalently%
\begin{equation}
v\left(  t\right)  =\sqrt{I_{B}}t~\ ,~\dot{u}=\frac{\left\vert \rho
_{m0}\right\vert }{6\sqrt{I_{B}}}t-\frac{\left(  V_{1}+V_{2}\right)  \left(
I_{B}\right)  ^{2}}{180}t^{6}.
\end{equation}
Notice, that the scale factor of the universe is approximated by
$a(t)=e^{u}\simeq e^{\frac{\left\vert \rho_{m0}\right\vert }{6\sqrt{I_{B}%
}t^{2}}}$ and $a(t)=e^{-\frac{\left(  V_{1}+V_{2}\right)  \left(
I_{B}\right)  ^{2}}{180}t^{7}}$ for small and large values of $t,$
respectively, where for the latter case we need to have $V_{1}+V_{2}<0$.
Furthermore, the expression for the Hubble function at small values of $t$ is
found to be%

\begin{equation}
H(a)\simeq a^{-12}(\ln a)^{1/2},
\end{equation}
while for large values of$~t$ we get%

\begin{equation}
H(a)\simeq-\left(  V_{1}+V_{2}\right)  a^{4}(\ln a)^{6/7} .
\end{equation}


\subsubsection{Radiation fluid}

As we have already discussed, Model B does not exist for $\gamma=\frac{4}{3}$
($w_{m}=1/3$). In this case the potential is
\begin{equation}
V\left(  \phi\right)  =V_{1}\phi^{2},
\end{equation}
while the corresponding conservation law generated by the same Killing Tensor
becomes%
\begin{equation}
I_{B}=a^{4}\dot{\phi}^{2}.
\end{equation}
Performing the transformation ~$a=r(t)[\cosh\vartheta(t)+\sinh\vartheta
(t)],~\phi=r(t)[\cosh\vartheta(t)-\sinh\vartheta(t)]$, we find that the
constraint equation and the conservation law $I_{B}^{\left(  r\right)  }$ are
written as
\begin{equation}
V_{1}r^{4}+6\dot{r}^{2}-2\rho_{0}+6r^{2}\dot{\vartheta}^{2}=0
\end{equation}

\begin{equation}
6r^{4}\dot{\vartheta}^{2}-I_{B}=0.
\end{equation}
from which we obtain directly:
\begin{equation}
\dot{\vartheta}^{2}=\frac{I_{B}}{6r^{4}}\ \ ,\ V_{1}r^{4}+6\dot{r}^{2}%
-2\rho_{0}+\frac{I_{B}}{r^{2}}=0.
\end{equation}
Solving the above set of equations, we can define the quantities $r(t)$ and
$\theta(t)$. For example, if we assume for simplicity that $\rho_{0}=I_{B}=0,$
then we find
\begin{equation}
r(t)=\pm\frac{6}{\sqrt{-6V_{1}}(c_{1}-t)}\ ,\vartheta(t)=c_{2},
\end{equation}
and thus $a\left(  t\right)  \simeq\frac{6}{\sqrt{-6V_{1}}(c_{1}-t)}%
,~\phi\left(  t\right)  \simeq\pm\frac{6}{\sqrt{-6V_{1}}(c_{1}-t)}.$ In this
case the Hubble function can now be directly calculated ($N=a$) and shown to
be a constant, that is $\left(  H(a)\right)  ^{2}=\frac{\dot{a}}{Na}%
=\frac{\left\vert V_{1}\right\vert }{6}$, which is the de Sitter solution.
That is an expected result since the current potential corresponds to the
$f\left(  R\right)  =R^{2}$ theory.

\subsection{Model C}

The third model that we find is that for which the constants $P$ and $Q$ are
expressed in terms of the barotropic index by%
\begin{equation}
P=\frac{3\gamma+5}{2}~,~Q=3\gamma.
\end{equation}
For this model the corresponding conservation law is calculated to be%
\begin{align}
I_{C}  &  =\left(  3\gamma-4\right)  a^{5-3\gamma}\phi^{2}\dot{a}^{2}+3\left(
\gamma-1\right)  a^{6-3\gamma}\phi\dot{a}\dot{\phi}+a^{7-3\gamma}\dot{\phi
}^{2}+\nonumber\\
&  ~~~~~~~\ \ -\frac{\left(  3\gamma-5\right)  }{6}V_{1}a^{3\gamma+1}%
\phi^{\frac{7+3\gamma}{2}}-\frac{3\gamma-5}{1+3\gamma}\gamma V_{2}\left(
a\phi\right)  ^{3\gamma+1}.
\end{align}
Using the transformation $a=z^{\frac{-2}{3\gamma-5}},\phi=wz^{\frac{2}%
{3\gamma-5}}$, we write the constraint equation and the conservation law in
the new coordinate system as
\begin{equation}
-V_{2}w^{3\gamma}-V_{1}\frac{w^{\frac{5+3\gamma}{2}}}{z}+\frac{5-3\gamma}%
{12z}p_{w}p_{z}-4\rho_{m0}=0,
\end{equation}

\begin{equation}
I_{C}=\frac{wp_{z}p_{w}}{72z}(5-3\gamma)^{2}\ +\frac{p_{z}^{2}\ }%
{144}(5-3\gamma)^{2}+\frac{V_{1}(5-3\gamma)}{6z}w^{\frac{7+3\gamma}{2}}%
+\frac{V_{2}\gamma(5-3\gamma)}{2(1+3\gamma)}w^{1+3\gamma},
\end{equation}
where now $\dot{z}=\frac{5-3\gamma}{12z}p_{w}~\ $and$~\dot{w}=\frac{5-3\gamma
}{12z}p_{z}.$

Following \cite{Daskaloy}, we can see that the dynamical system is supported
by a Lie surface and the solution of the Hamilton-Jacobi equation provides the
action%
\begin{align}
S\left(  z,w\right)   &  =z\sqrt{Ic+\frac{24V_{2}}{(5-3\gamma)(3\gamma
+1)}w^{3\gamma+1}+\frac{48\rho_{m0}}{5-3\gamma}w}\\
&  +\ \int dw\frac{\frac{12V_{1}}{5-3\gamma}w^{\frac{5+3\gamma}{2}}}%
{\sqrt{Ic+\frac{24V_{2}}{(5-3\gamma)(3\gamma+1)}w^{3\gamma+1}+\frac
{48\rho_{m0}}{5-3\gamma}w}},
\end{align}
where $\gamma\neq5/3$.

\subsubsection{Special case $\gamma=5/3$}

Again, in the special case of $\gamma=\frac{5}{3}$ we utilize the coordinates
of section 3.1.1 $\left(  u,v\right)  $. Therefore, the constraint equation
and the conservation law $I_{c}^{\left(  u\right)  }$ now take the form
\begin{equation}
-p_{u}p_{v}+6(V_{1}+V_{2})v^{5}+4\rho_{m0}=0,
\end{equation}
$\ $%
\begin{equation}
\frac{p_{u}^{2}}{36}-I_{C}=0.
\end{equation}
Now, we obtain
\begin{equation}
\dot{v}=\sqrt{I_{c}^{\left(  u\right)  }}~\text{and }\dot{u}=\frac{\rho_{m0}%
}{9\sqrt{I_{c}^{\left(  u\right)  }}}+\frac{\left(  V_{1}+V_{2}\right)
}{6\sqrt{I_{c}^{\left(  u\right)  }}}v^{5},
\end{equation}
where $\dot{v}=\frac{1}{6}p_{u}~,~\dot{u}=\frac{1}{6}p_{v}$. Hence, we can
write an analytic solution in closed-form, namely
\begin{equation}
v(t)=\sqrt{I_{C}}t~\text{and }u(t)=\frac{\rho_{m0}}{9\sqrt{I_{C}}}%
t+\frac{\left(  V_{1}+V_{2}\right)  (I_{C})^{\frac{9}{2}}}{36}t^{6}+c,
\end{equation}
and thus we obtain%

\begin{equation}
a(t)=\exp\left(  \frac{\rho_{m0}}{9\sqrt{I_{C}}}t+\frac{\left(  V_{1}%
+V_{2}\right)  (I_{C})^{\frac{9}{2}}}{36}t^{6}\right)  +c.
\end{equation}
For small values of $t$ we easily see that $a(t)\simeq\exp\left(
A_{0}t\right)  +c$, while for large values of $t$ the scale factor becomes
$a(t)\simeq\exp\left(  A_{1}t^{6}\right)  +c$. Using the expression of the
Hubble function $H(a)=\frac{1}{N}\frac{\dot{a}}{a}$ where now $N=a^{2}$ we find%

\begin{equation}
H(a)\simeq a^{-\frac{3}{4}}-ca^{-\frac{7}{4}}%
\end{equation}
for small values of $t,$ and
\begin{equation}
H(a)\simeq6\left(  A_{1}\right)  ^{\frac{1}{6}}\left(  \ln\left(  a-c\right)
\right)  ^{\frac{5}{6}}\left(  a-c\right)  a^{-\frac{7}{4}},
\end{equation}
for large values of the cosmic time. Finally, in this scenario the
corresponding action is given by%

\begin{equation}
S\left(  u,v\right)  =I_{C}^{-\frac{1}{2}}\left[  (V_{1}+V_{2})v^{6}%
+24v\rho_{m0}+36uI_{C}\right]  .
\end{equation}

\section{Quintessence and acceleration for empty spacetime}

\label{quint}

The action integral of the O'Hanlon field $\phi~$in a 4-dimensional empty
spacetime is given from the following expression:
\begin{equation}
S=\int dx^{4}\sqrt{-g}\left[  \phi R-V\left(  \phi\right)  \right]  .
\label{CLN.12}%
\end{equation}
Under the conformal transformation $\bar{g}_{ij}$ $=\phi g_{ij}$
Eq.(\ref{CLN.12}) becomes%
\begin{equation}
S_{MN}=\int dx^{4}\sqrt{-\bar{g}}\left[  -\frac{\bar{R}}{2}+\frac{1}{2}\bar
{g}^{ij}\nabla_{\mu}\psi\nabla_{\nu}\psi-e^{-\frac{2\psi}{\sqrt{3}}}V\left(
\psi\right)  \right]  , \label{CLN.12.7}%
\end{equation}
where $\bar{R}$ now is the Ricci scalar of the conformal metric $\bar{g}_{ij}$
and $\psi$ is related with the field $\phi$ as
\begin{equation}
\psi=\sqrt{3}\ln\phi\text{ or }\phi=\exp\left(  \frac{\sqrt{3}}{3}\psi\right)
. \label{CLN.12.8}%
\end{equation}
The main difference between the fields $\phi$ and$~\psi$ is that they are
defined in different frames. Specifically, $\psi$ is minimally coupled scalar
field, while $\phi$ is non minimally coupled and defined in the Jordan frame.

In our case, with the aid of Eq.(\ref{CLN.12.8}), the family of potentials
provided by Eq.(\ref{ac.21}) is written in terms of $\psi$ as follows
\begin{equation}
\bar{V}\left(  \psi\right)  =V_{1}\exp\left(  \alpha\psi\right)  +V_{2}%
\exp\left(  \alpha\psi\right)  , \label{CLN.12.9}%
\end{equation}
in which $\alpha=\left(  P-2\right)  /\sqrt{3}$ and $\beta=\left(  Q-2\right)
/\sqrt{3}$.

This class of potentials has been studied in the early universe in ref.
\cite{barreiro}, and recently a special case of this family of potentials has
been investigated in the constant-roll inflationary regime in ref.
\cite{starin}. Barreiro et al. \cite{barreiro} have shown that
Eq.(\ref{CLN.12.9}) can describe various scaling solutions in the early
universe which depend on the constants appearing in the exponential terms,
namely $\alpha$ and $\beta=\frac{\left(  Q-2\right)  }{\sqrt{3}}$.\emph{ }As
was discussed in \cite{barreiro}, one of the coefficients $\alpha,~\beta$ has
to be greater than $5.5$, (for example, we assume $\left\vert \alpha
\right\vert >5.5$)$,$ in order for the theory to be in agreement with the
constraints imposed by the primordial nucleosynthesis. Furthermore, in order
to have a quintessence description with equation of stater parameter that fits
observations, the second free parameter is constrained by\ $\beta<0.8$.
Moreover, for those values of the parameters where $\alpha\beta>0,~$there
exists a late-time attractor which describes and accelerated universe with
equation of state parameter different from $-1$. However, for $\alpha\beta<0$
the equation of state parameter of the late-time attractor has an oscillatory
behavior until it mimics the cosmological constant.

\begin{figure}[t]
\centering\includegraphics[scale=0.45]{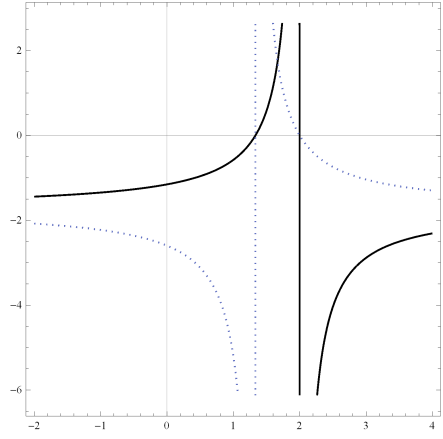}
\caption{Exponential rates of the early universe potential (\ref{CLN.12.9})
for the family class of models B which corresponds to the Ermakov-Pinney
system.}%
\label{plot011}%
\end{figure}

In Fig. \ref{plot011} we present the exponential rates of the potential
(\ref{CLN.12.9}) for the second family class of models. If we assume that
$\gamma$ is to be the parameter of a barotropic fluid, that is, $\gamma
\in\lbrack1,2)~$the the vertical axis is constrained by the same bounds. In
general, mathematically $\gamma$ can take any value except $\gamma\neq\frac
{4}{3}.$ However, close to the value $\gamma=\frac{4}{3}$, which corresponds
to the limit where the matter source is radiation, we can see that our
parameters are in agreements with the bounds that we discussed on the previous
paragraph and specifically those arising when $\alpha\beta>0.$

\section{Conclusions}

\label{con}

In this work, we have provided new analytical solutions for $f(R)$ gravity
models which contain an ideal fluid in a spatially-flat FLRW universe. In
order to determine the $f\left(  R\right)  $ models which admit exact
solutions we applied the well known method of invariant transformations and,
with the aid of Noether's second theorem, we constructed the corresponding
conservation law which was used to solve the Hamilton-Jacobi equation. Then,
this solution was applied in order to reduce the field equations to a system
of two first-order ordinary differential equations and express the solution in
closed-form (where possible). It is interesting to mention that some of the
current $f(R)$ models have been discussed in Ref. \cite{rr1}, and in the
framework of Brans-Dicke gravity in Ref.\cite{and2}. However, in this work we
found new cases of integrable $f(R)$ models. Therefore, in the context of
spatially flat FLRW metrics, the combination of the present work with those of
Paliathanasis et al. \cite{rr1} and Papagianopoulos et al. \cite{and2}
provides a complete study of the integrable $f(R)$ models that contain
conservation laws that are quadratic in momentum.
Finally, since the issue of frames are still open, namely if analytical
solutions have to be interpreted either in the Einstein frame or in the Jordan
frame, we have studied conformally related metrics and scalar fields in the
framework of O'Hanlon's theory. In the Jordan frame, we determined the
corresponding minimally coupled scalar field potential which corresponds to
the integrable models that we found in the Einstein frame.\emph{ }

\begin{acknowledgments}
S. Basilakos acknowledges support by the Research Center for Astronomy of the
Academy of Athens in the context of the program \textit{Testing general
relativity on cosmological scales} (ref. number 200/872). AP was financially
supported by FONDECYT grants 3160121. AP thanks the University of Athens for
the hospitality provided while part of this work was performed. J.D. Barrow
acknowledges support from the Science and Technology Facilities Council (STFC)
of the UK.
\end{acknowledgments}

%

\appendix

\section{Symmetries and Conservation laws}

\label{LBsym}

In this Appendix we present the basic mathematical material which applied in
this work towards determining the conservation law of the gravitational field equations.

Consider a differential function $F\left(  x^{i},u^{A},u_{,i}^{A},u_{,ij}%
^{A}...\right)  $ which is defined in the jet space$\left\{  x^{i}%
,u^{A},u_{,i}^{A},u_{,ij}^{A},...\right\}  $ where $x^{i}$ denotes the
independent variables and $u^{A}$ are dependent variables. \ The differential
function $F$ will be invariant under the infinitesimal transformation%
\begin{align}
\bar{x}^{i}  &  =x^{i}+\varepsilon\xi^{i}\left(  x^{i},u^{B},u_{,i}%
^{B},u_{,ij}^{B}...\right)  ,\label{LB.01}\\
\bar{u}^{A}  &  =u^{A}+\varepsilon\eta^{A}\left(  x^{i},u^{B},u_{,i}%
^{B},u_{,ij}^{B}...\right)  , \label{LB.02}%
\end{align}
if and only if $F\left(  x^{i},u^{A},u_{,i}^{A},u_{,ij}^{A}...\right)
=\bar{F}\left(  \bar{x}^{\bar{\imath}},\bar{u}^{\bar{A}},\bar{u}_{,\bar
{\imath}}^{\bar{A}},\bar{u}_{,\bar{\imath}\bar{j}}^{\bar{A}}...\right)
,~$which means that at every point the value of the differential function will
be the same \cite{Sarlet,Crampin}. This is equivalent to the following
mathematical expression
\begin{equation}
X^{\left[  n\right]  }F=\lambda F, \label{LB.02a}%
\end{equation}
where $X$ is the generator of infinitesimal transformation (\ref{LB.01}%
)-(\ref{LB.02}) defined as
\begin{equation}
X=\xi^{i}\left(  x^{i},u^{B},u_{,i}^{B},u_{,ij}^{B}...\right)  \partial
_{i}+\eta^{A}\left(  x^{i},u^{B},u_{,i}^{B},u_{,ij}^{B}...\right)
\partial_{u}. \label{LB.03}%
\end{equation}
$X^{\left[  n\right]  }$ denotes its extension in the jet space $\left\{
x^{i},u^{A},u_{,i}^{A},u_{,ij}^{A},...\right\}  ~$and $\lambda=\lambda\left(
x^{i},u,u_{,i},u_{,ij}...\right)  $ is a function which should determined. If
there is a function $\lambda$ such that the condition (\ref{LB.02a}) holds
then the generator $X$ is called a symmetry of the differential function.

The functional form of the generator (\ref{LB.03}) defines the kind of
symmetries. For instance, when $\xi^{i},\eta^{A}$ are functions only of
$\left\{  x^{i},u^{A}\right\}  $ then $X$ is called a point symmetry, while if
$\xi^{i},\eta^{A}$ are linear in the first derivatives, $u_{,i}^{B},$ then $X$
is called contact symmetry. Of course, all the vector fields are Lie
symmetries because we are dealing with local infinitesimal transformations.

There are various methods to construct conservation laws for differential
functions/equations with the aid of the symmetry vectors. However, the
simplest method is to apply Noether's theorems. The first of these states that
if the action integral is invariant under the action of an infinitesimal
transformation then the field equations are invariant. For lagrangian
functions of the form $\mathcal{L}\left(  t,q^{i},\dot{q}^{i}\right)  ,$ which
describe second-order differential equations, Noether's first theorem takes
the form,%
\begin{equation}
X^{\left[  1\right]  }\mathcal{L}+\mathcal{L}\frac{d\xi}{dt}=\Phi,
\label{LB.04}%
\end{equation}
where $\xi$ is the component of the generator $X$ in the direction of the
independent variables and $\Phi=\Phi\left(  t,q,\dot{q}\right)  $ is a
boundary function.

In our discussion the lagrangian of the field equations has the form
$\mathcal{L}=T-V$; $\ $that is, $\mathcal{L}\left(  q^{k},\dot{q}^{k}\right)
=\frac{1}{2}\gamma_{ij}\dot{q}^{i}\dot{q}^{j}-V\left(  q^{k}\right)  $ where
$\gamma_{ij}\left(  q^{k}\right)  $ is the minisuperspace metric. Therefore,
with this family of lagrangians, and for contact symmetries in which
$X=K_{j}^{i}\left(  t,q^{k}\right)  \dot{q}^{i}\partial_{i}$, the Noether
symmetry condition (\ref{LB.04}) gives the condition%
\begin{equation}
K^{ij}V_{j}+\Phi_{,i}=0,
\end{equation}
where $K_{ij}=K_{ij}\left(  q^{k}\right)  $ is a Killing tensor of the minisuperspace.

Finally, Noether's second theorem can be applied to write the explicit form of
the corresponding conservation law, which for these types of contact symmetry
takes the simple form, $I=K_{ij}\dot{q}^{j}\frac{\partial\mathcal{L}}%
{\partial\dot{q}^{i}}-\Phi.$

\end{document}